\documentclass[12pt,a4paper]{article}
\usepackage[affil-it]{authblk}

\usepackage{amsmath,dsfont,bbold,slashed,tensor,amssymb,epsf,epsfig,amsthm,bm,graphicx}
\usepackage{scalerel}
\usepackage{gensymb}
\usepackage{subfig}
\usepackage[font=scriptsize,labelfont=bf]{caption}
\usepackage{float}
\usepackage{doi}
\usepackage{hyperref}
\usepackage{cite}
\usepackage[noabbrev]{cleveref}
\usepackage{color}
\usepackage{comment}
\usepackage[top=2.2cm, bottom=2.2cm, left=2cm, right=2cm]{geometry}

\numberwithin{equation}{section}

\definecolor{darkgreen}{rgb}{0,0.35,0}

\newcommand\beq{\begin{equation}}
\newcommand\eeq{\end{equation}}
\newcommand\bal{\begin{aligned}}
\newcommand\eal{\end{aligned}}

\newcommand{\UACh}{Instituto de Ciencias F\'isicas y Matem\'aticas, Universidad Austral de Chile, Casilla 567, 5090000, Valdivia, Chile}
\newcommand{\UCharles}{IPNP - Faculty of Mathematics and Physics, Charles University, V Hole\v{s}ovi\v{c}k\'ach 2, 18000 Prague 8, Czech Republic}
\newcommand{\WroUT}{Institute for Theoretical Physics, Wroc\l{}aw University of Science and Technology, 50-370 Wroc\l{}aw, Poland}
\newcommand{\UNAPa}{Instituto de Ciencias Exactas y Naturales, Universidad Arturo Prat, Playa Brava 3265, 1111346, Iquique, Chile}
\newcommand{\UNAPb}{Facultad de Ciencias, Universidad Arturo Prat, Avenida Arturo Prat Chac\'on 2120, 1110939, Iquique, Chile}

\begin{document}

\title{\textbf{A note on the Hamiltonian structure of transgression forms}\vskip1truecm}

\author[1,2]{Pablo Pais\thanks{pais@ipnp.troja.mff.cuni.cz}}
\affil[1]{\UACh}
\affil[2]{\UCharles}

\author[3]{Patricio Salgado-Rebolledo\thanks{patricio.salgado-rebolledo@pwr.edu.pl}}
\affil[3]{\WroUT}

\author[4,5]{Aldo Vera\thanks{aldoveraseron@gmail.com}}
\affil[4]{\UNAPa}
\affil[5]{\UNAPb}

\date{\vskip1.5truecm}

\maketitle

\begin{abstract}
By incorporating two gauge connections, transgression forms provide a generalization of Chern-Simons actions that are genuinely gauge-invariant on bounded manifolds. In this work, we show that, when defined on a manifold with a boundary, the Hamiltonian formulation of a transgression field theory can be consistently carried out without the need to implement regularizing boundary terms at the level of first-class constraints. By considering boundary variations of the relevant functionals in the Poisson brackets, the surface integral in the very definition of a transgression action can be translated into boundary contributions in the generators of gauge transformations and diffeomorphisms. This prescription systematically leads to the corresponding surface charges of the theory, reducing to the general expression for conserved charges in (higher-dimensional) Chern-Simons theories when one of the gauge connections in the transgression form is set to zero.
\end{abstract}

\newpage
\tableofcontents


\section{Introduction}

Chern-Simons (CS) forms have been extensively used in the construction of action principles in different areas of physics (see \cite{Dunne:1998qy,Deser:1998mk,Deser:1998it,marino2005chern,Zanelli:2008sn,fradkin2013field} and references therein). In the context of gravity, CS forms can be used to define (super-)gravity theories in odd dimensions as genuine gauge theories \cite{Achucarro:1987vz,Witten:1988hc,Chamseddine:1989nu,Troncoso:1998ng} (for a detailed analysis, see also \cite{Zanelli:2005sa,hassaine2016chern}). However, when gauge fields are defined on a manifold with a boundary, CS actions become \emph{quasi-gauge-invariant}. In other words, under gauge transformations, they change in an exact form. In order to restore gauge invariance, boundary terms can be added to the action (see, for instance, \cite{Geiller:2017xad,Arcioni:2002vv}). Furthermore, boundary terms together with boundary conditions on the fields are generally necessary for the action to have a well-defined variation and be a true extremum when the field equations hold \cite{Regge:1974zd}. In the case of three-dimensional CS gravity with a negative cosmological constant, this procedure generally leads to a (reduced) Wess-Zumino-Witten (WZW) model defined at the two-dimensional boundary \cite{Coussaert:1995zp}. These results have been generalized to the five-dimensional case \cite{Gegenberg:1999zu}, where a WZW$_4$ model is shown to arise at the boundary when suitable boundary conditions are adopted. The study of gauge theories defined on bounded manifolds has attracted considerable attention in the last decades due to the advent of the holographic principle \cite{tHooft:1993dmi,Susskind:1994vu} and the AdS/CFT correspondence \cite{Maldacena:1997re}. In this context, bulk/boundary dualities in CS theories have led to remarkable results in the direction of a holographic description of three-dimensional gravity (see, for example, \cite{Brown:1986nw,Strominger:1997eq,Gukov:2004id,Witten:2007kt}). Moreover, in recent years, different aspects of dual field theories associated to higher-dimensional AdS CS gravities have been explored \cite{Banados:2004zt,Banados:2005rz,Cvetkovic:2017fxa}.

When applying Dirac's Hamiltonian formalism for constrained systems \cite{Dirac,Henneaux-Teitelboim} to CS theories in the presence of a boundary, conserved charges appear as surface integrals that, provided a set of boundary conditions, regularize the first-class constraints of the system and allow one to find a surface charge algebra once the gauge is appropriately fixed and the Dirac brackets are implemented. This is nothing but a particular example of the prescription due to Regge and Teiteilboim \cite{Regge:1974zd} applied to gauge theories. Moreover, in $2+1$ dimensions, the Hamiltonian formalism for CS theories is \emph{generic}, i.e. the rank of the symplectic form that defines the phase space of the theory is constant, and the corresponding Dirac brackets can be globally defined \cite{Banados:1994tn,Park:1998yw}. After imposing suitable boundary conditions on the CS gauge connection, one can find the surface charge algebra of the theory from the Dirac brackets of the regularized first-class constraints.  On the other hand, in higher dimensions, the situation changes completely, and the  Hamiltonian dynamics of the theory is highly non-trivial.  This is because, for $D\geq5$, the CS symplectic form has a non-constant rank throughout the phase space, dividing this into \emph{sectors} possessing different numbers of local degrees of freedom. The Hamiltonian formulation, in this case, can be used to find the conserved charges once a generic and regular sector of the theory has been chosen, and the constraints of the theory can be adequately split into first- and second-class \cite{Banados:1996yj,Miskovic:2005di}.

Transgression forms first appeared in physics in the topological interpretation of anomalies in quantum field theory and gravity \cite{Alvarez-Gaume:1984zlq,Manes:1985df}. In the context of gravitational gauge theories, transgression field theory (TrFT) can be used to define (super-)gravity action principles in odd dimensions that are genuinely gauge invariant \cite{Mora:2004kb,Izaurieta:2006aj,Mora:2006ka,Mora:2014aca}, whereas, in  the even-dimensional case, TrFTs allow one to define gauged-WZW models and topological gravity theories \cite{Anabalon:2006fj,Anabalon:2007dr,Mora:2011sz,Merino:2010zz,Salgado:2013pva,Salgado:2014jka}. Being genuinely gauge-invariant, it is possible to define conserved charges in TrFTs through Noether’s theorem straightforwardly. Transgression forms can be constructed by subtracting two CS forms plus a boundary term, rendering the theory gauge invariant. Therefore, it is reasonable to expect that in the case of a TrFT, the boundary term in its very definition is sufficient to obtain the corresponding conserved charges as surface integrals in the symmetry generators through Dirac's algorithm. In this article, we will show that, provided the boundary variations of the relevant functionals are consistently treated, this is indeed the case.

As stated before, the usual criteria when applying the Regge-Teitelboim method to gauge theories is to include regularizing boundary terms in the first-class constraints \cite{Benguria:1976in,Henneaux:1985tv,Brown:1986nw} so that their variations do not drop boundary terms and their Poisson brackets lead to regularized functionals as well \cite{Brown:1986ed}. However, to apply the Hamiltonian formulation for constrained systems to an action based on a transgression form and not to lose the information contained in the transgression boundary term during the process, one should necessarily allow boundary contributions in the variations of the constraints of the theory. As shown by Soloviev and Bering \cite{Soloviev:1993zf,Bering:1998fm} (see also \cite{Soloviev:1998nm}), treating functionals with boundary variations  requires carefully extending the definition of the usual Poisson bracket to the boundary of the space-time manifold.\footnote{Another way to circumvent this problem has been developed in \cite{BarberoG.:2019rfb} by means of a purely geometric formulation of Dirac's formalism for constrained systems in bounded regions, which does not require implementing regularizing terms in the constraints.} This is mainly due to the fact that when boundary variations are allowed, the Poisson bracket does not fulfil the Jacobi identity \cite{Soloviev:1992be}. Based on these results, we will analyze TrFT applying Dirac's formalism for constrained systems and including the boundary variations of the phase space functionals throughout. The inclusion of boundary variations in the Poisson brackets effectively leads to a Hamiltonian formulation based on the modified Poisson bracket proposed by Soloviev and Bering \cite{Soloviev:1993zf,Bering:1998fm}. We will show that this procedure allows us to recover the conserved charges of TrFT from the resulting boundary terms in the first-class constraints of the theory.

This paper is organized as follows. \Cref{transgression} will briefly review the definition of transgression forms and their relation to CS theory. In \Cref{hamiltonianformulation}, we give a short summary of the Hamiltonian formulation of CS theories. We introduce a generalized Poisson bracket and use it to apply Dirac's algorithm for constrained systems to TrFT. In \Cref{3y5D}, we apply this prescription to transgression forms in three and five dimensions and show that the boundary terms obtained in the first-class constraints allow us to recover the conserved charges of the theory. In \Cref{conclusions}, we conclude with an analysis of the results and discuss possible future directions.

\section{Transgression Forms}
\label{transgression}

We start by briefly surveying the properties of transgression forms and TrFT. This summary makes no pretence of being complete or up-to-date. More detailed analysis can be found in the references (see, for instance, \cite{bertlmann2000anomalies,Nakahara}).

Given a principal bundle with fibre $G$ over a $(2n+2)$-dimensional base manifold and a connection one-form $A$ defined on it, one can always construct an invariant polynomial $P(F)$ out of the $(n+1)$-th wedge power of the curvature two-form $F=dA+A^2$ associated with $A$, which is usually expressed as\footnote{Notice that wedge product between differential forms is understood.}
\beq\label{invariant_polynomial}
P(F)=\langle F\ldots F\rangle\;,
\eeq
where $\langle\dots\rangle$ stands for an invariant symmetric multilinear form on the Lie algebra $\mathfrak{g}$ associated with the Lie group $G$. By definition, the polynomial \eqref{invariant_polynomial} is a $(2n+2)$-form invariant under gauge transformations $F\rightarrow g^{-1}Fg$ where $g\in G$. The Chern-Weil theorem \cite{Nakahara} ensures that $P(F)$ is closed [$dP(F)=0$] and that, if $F$ and $\bar{F}$ are the curvatures corresponding to two different connections $A$ and $\bar{A}$ on the same principal bundle, $P(F)-P(\bar{F})$ is exact, i.e.
\beq\label{chernweyl}
P(F)-P(\bar{F})=d\mathcal{T}_{2n+1} \ .
\eeq
This defines the transgression form
\beq
\mathcal{T}_{2n+1}\left[A,\bar{A}\right]=(n+1)\int_{0}^{1}dt\ \left\langle \left({A-}\bar{A}\right)F_{t}^{n}\right\rangle\;, \label{transgression_definition}
\eeq
where $t\in [0,1]$, $A_{t}=tA+(1-t)\bar{A}$ is a connection that interpolates between $A$ and $\bar{A}$, and $F_{t}=dA_{t}+A_{t}^{2}$ is the corresponding interpolating curvature \cite{Alvarez-Gaume:1984zlq,Manes:1985df} (for details on the mathematical definition of the transgression see \cite{bertlmann2000anomalies,Nakahara}). From \cref{transgression_definition}, it is straightforward to show that the transgression form is gauge-invariant.
It is important to note that setting $\bar{A}=0$ leads to the CS $(2n+1)$-form
\beq\label{CSa}
CS_{2n+1}\left[A\right]=\mathcal{T}_{2n+1}\left[A,0\right]=(n+1)\int_{0}^{1}dt\ \left\langle A \left(tdA+t^2 A^2 \right)^{n}\right\rangle\;.
\eeq
Furthermore, the transgression form \eqref{transgression_definition} can be written as the difference of two CS forms plus an exact term
\beq\label{transgression_definition_2}
\mathcal{T}_{2n+1}\left[A,\bar{A}\right]=CS_{2n+1}\left[A\right]-CS_{2n+1}\left[\bar{A}\right]-d\mathcal{B}_{2n}\left[A,\bar{A}\right]\;,
\eeq
where $\mathcal{B}_{2n}$ is given by
\beq\label{Bterm}
\mathcal{B}_{2n}= -n(n+1)\int_0^1 ds \int_0^1 dt s \left\langle A_t (A-\bar{A}) F_{st}^{n-1}  \right\rangle , \quad F_{st}=sF_t+s(s-1)A_t^2\;.
\eeq
At this point it is important to remember that the CS form \eqref{CSa} is not strictly gauge-invariant, but \emph{quasi-invariant}. Indeed, its variation under gauge transformations of the gauge connections $A$ and $\bar{A}$ drops an exact form. It is precisely the boundary term in \cref{transgression_definition_2} that makes the transgression a truly gauge-invariant form. Due to this property, transgression forms have been used to define gauge-invariant actions for field theories and gravity in diverse dimensions \cite{Mora:2004kb,Izaurieta:2006aj,Mora:2006ka}.
\begin{figure}
\centering
\includegraphics[width=.5\textwidth,angle=0]{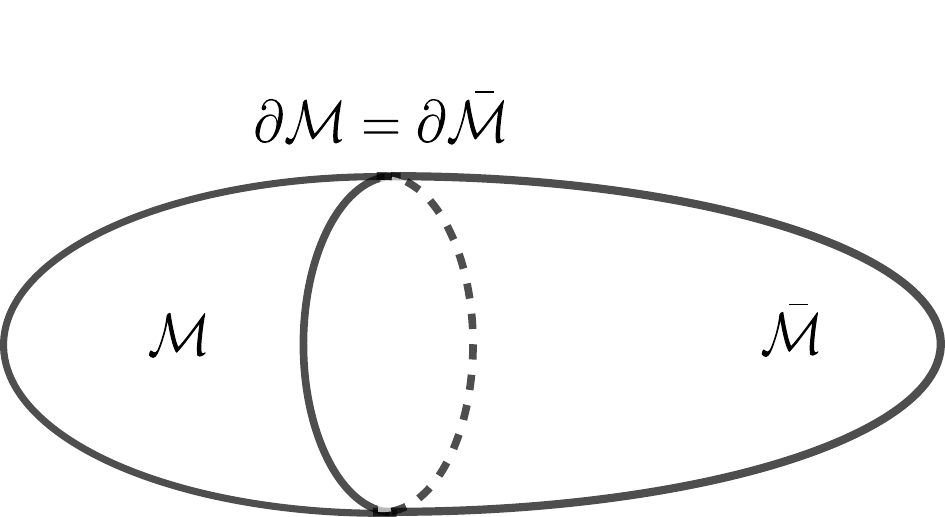}
  \caption{\textbf{Cobordant manifolds.} Two manifolds $\mathcal{M}$ and $\bar{\mathcal{M}}$ having a common boundary $\partial \mathcal{M}=\partial\bar{\mathcal M}$. }\label{Figure_cobordism}
\end{figure}

A TrFT is a field theory whose action is constructed out of a transgression form. Different variants of TrFT can be defined depending on the way the gauge connections $A$ and $\bar A$ are treated. For instance, one could consider $A$ and $\bar A$ as respectively defined on two different $(2n+1)$-dimensional manifolds, $\mathcal M$ and $\bar{\mathcal M}$, that are \textit{cobordant}, i.e. $\partial \mathcal{M}=\partial\bar{\mathcal M}$ (see \Cref{Figure_cobordism}). The principal bundle with fiber $G$ can then be defined on the manifold $\mathcal M+\bar{\mathcal M}$ and the definition \eqref{transgression_definition_2} can be used to define the following action principle for a TrFT \cite{Mora:2006ka}
\beq\label{action}
\begin{aligned}
I\left[A,\bar{A}\right] =\kappa\int_{\mathcal{M}}CS_{2n+1}\left[A\right]-\kappa\int_{\mathcal{\bar{M}}}CS_{2n+1}\left[\bar{A}\right]-\kappa\int_{\partial\mathcal{M}}\mathcal{B}_{2n}\left[A,\bar{A}\right] \ ,
\end{aligned}
\eeq
where $\kappa$ is a constant.\footnote{Notice that when using TrFT to define a generalization of CS gravity \cite{Mora:2006ka}, $\kappa$ is defined in terms of the $D$-dimensional Newton constant $G$ as $\kappa=[2(d-2)!\Omega_{d-2}G]^{-1}$ (with $\Omega_d$ the volume of the sphere $S^d$), and is usually absorbed in the definition of the invariant multilinear form introduced in  \cref{invariant_polynomial}.} In general, one can allow the connections $A$ and $\bar{A}$ to transform with different group elements, provided they match at the boundary, i.e. the action \eqref{action} is invariant under transformations of the form
\beq
A \longrightarrow gAg^{-1}-dgg^{-1}\ , \quad  \bar{A} \longrightarrow \bar{g}A\bar{g}^{-1}-d\bar{g}\bar{g}^{-1} \ , \quad g\Big|_{\partial\Sigma}=\bar{g}\Big|_{\partial\Sigma} \ .
\eeq
In the case of infinitesimal gauge transformations, where $g=\mathbb{1}+\eta$, $\bar{g}=\mathbb{1}+\bar{\eta}$, and $\eta,\ \bar{\eta}\,\in\mathfrak{g}$, this means
\begin{align}   \label{gauge_transf_infinitesimal}
\delta A = -D\eta \ , \quad  \delta\bar{A}=-\bar{D}\bar{\eta} \ , \quad \eta\Big|_{\partial\Sigma} = \bar{\eta}\Big|_{\partial\Sigma}\;.
\end{align}
The action \eqref{action} has been used to generalize CS gravitational theories, providing the conserved charges of the theory and the correct expression for black hole thermodynamics when suitable boundary conditions are adopted \cite{Mora:2004kb,Mora:2006ka,Aros:2006ke}.

\begin{figure}
  \centering
  \includegraphics[width=0.6\textwidth,angle=0]{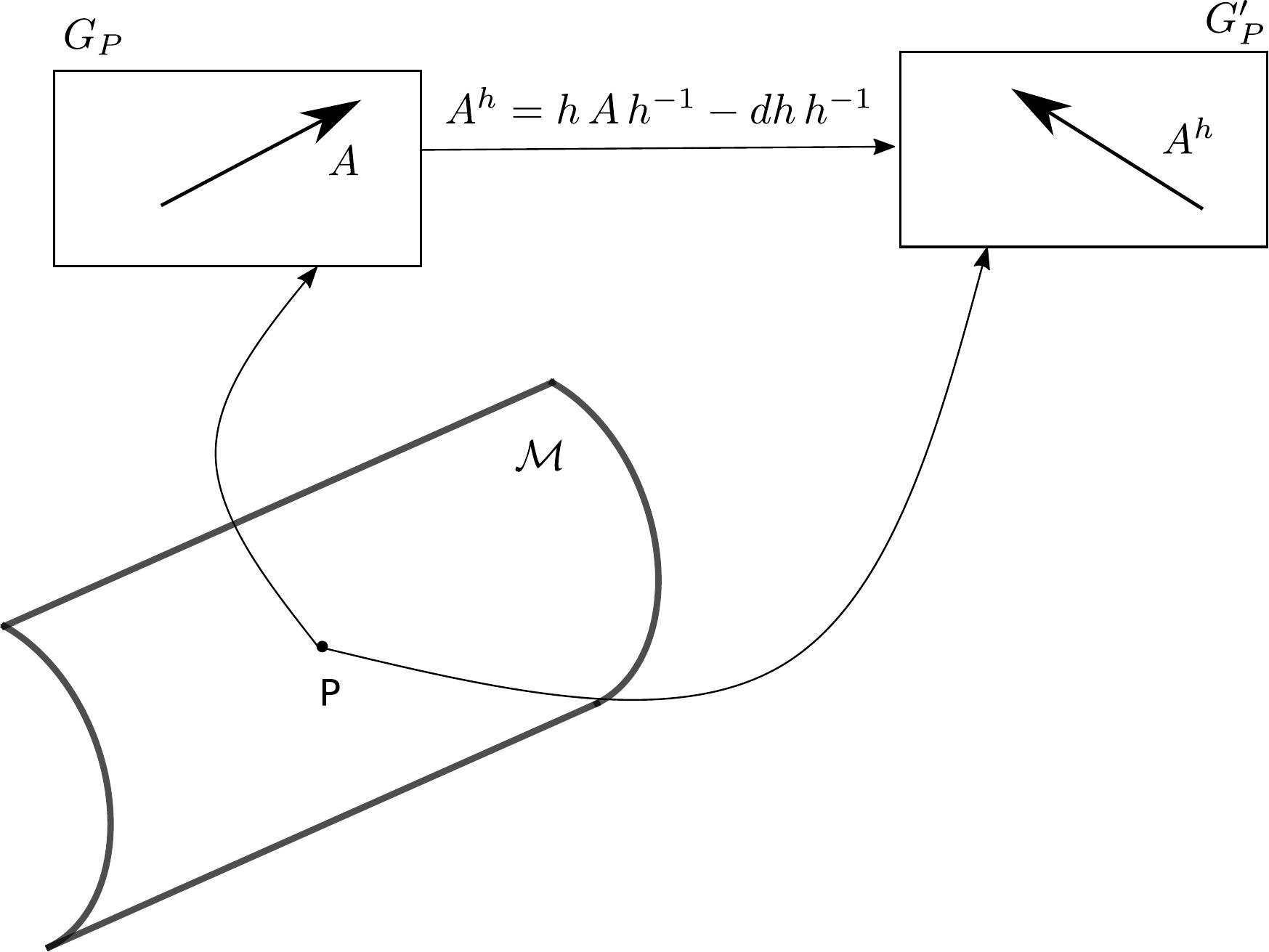}
  \caption{\textbf{Gauged WZW case.} A single manifold $\mathcal{M}$ where two connections $A$ and $A^{h}$ are defined, belonging to their respective spaces $G_{p}$ and $G'_{p}$ at $p\in\mathcal{M}$. $A$ and $A^{h}$ are connected by a gauge transformation defined by an element $h\in G$, where $G$ is the strucutre group.}\label{Figure_non_cobordism}
\end{figure}

Another type of TrFT can be constructed by considering both connections, $A$ and $\bar A$, as defined on the same base manifold, i.e. $\mathcal{M}=\bar{\mathcal M}$. In particular, one can consider $A$ and $\bar A$ as belonging to the same gauge orbit, $\bar{A}=A^h=hAh^{-1}-dh h^{-1}$,
where $h\in G$ (see \Cref{Figure_non_cobordism}). In this case, a TrFT reduces to a gauged WZW model living at the boundary \cite{Anabalon:2006fj,Mora:2011sz,Salgado:2013pva,Salgado:2014jka}.

In this work, we will consider the connections $A$ and $\bar A$ as independent, but defined on the same manifold $\mathcal M$. However, the results can be easily generalized to the case of cobordant manifolds discussed above. Considering $\mathcal M=\bar{\mathcal M}$ in \cref{action} allows one to write the TrFT action as simply the integral of the transgression form \eqref{transgression_definition_2} over $\mathcal M$,
\beq\label{action2}
I\left[A,\bar{A}\right] =
\kappa\int_{\mathcal M} \mathcal{T}_{2n+1}\left[A,\bar{A}\right]\;.
\eeq

\section{Hamiltonian structure of transgression field theory}
\label{hamiltonianformulation}

In this section, we apply Dirac's formalism for constrained systems to the TrFT defined by the action \eqref{action2}. Since the boundary term in the definition of the transgression form guarantees the gauge invariance of the theory, we will consider all the contributions coming from this term throughout the Hamiltonian formalism and will not add any regularizing boundary term to the constraints by hand. This procedure can be implemented by defining a generalized Poisson bracket  that takes into account the boundary variations of the phase space functionals. This bracket was introduced by Soloviev and Bering in references \cite{Soloviev:1993zf} and \cite{Bering:1998fm}. By means of this generalized bracket, the boundary terms present in the transgression action will be shown to lead to surface integrals for the first-class constraints of the theory. The generators of the symmetry transformations constructed in terms of these first-class constraints form a Poisson algebra isomorphic to the corresponding symmetry algebras of the theory under the modified Poisson bracket. Furthermore, its surface integrals provide a general formula for the conserved charges of the theory. The important point in this derivation is that it does not need to adopt any ``a priori'' boundary conditions on the fields or regularizing boundary terms.

\subsection{Review of Chern-Simons theories}
\label{CSanalysis}

Let us start reviewing the Hamiltonian formulation of (higher-dimensional) CS theories. We will  consider a $(2n+1)$-dimensional space-time without boundary and with topology $\mathcal{M}=\mathbb{R}\times\Sigma$, where $\mathbb{R}$ corresponds to the real (temporal) line and $\Sigma$ is a $2n$-dimensional spatial hypersurface. We denote the infinitesimal generators of $G$ as $T_a$, which satisfy
\beq\label{LieAlg}
[\![T_a, T_b]\!]= f^c_{ab} T_c \ ,
\qquad
\left\langle T_{a_1}\cdots T_{a_{n+1}}\right\rangle=g_{a_1\cdots a_{n+1} }\;,
\quad
a=1,\dots,{\rm dim}\,\mathfrak g\;,
\eeq
with $f^c_{ab}$ the structure constants of the Lie algebra $\mathfrak g$ ang $g_{a_1\cdots a_{n+1} }$ an invariant symmetric $(n+1)$-rank tensor. In this case, a connection one-form taking values on $\mathfrak g$ can be split as\footnote{Greek indices $\mu,\nu,\rho,\ldots$ denote space-time directions and range from $0$ to $2n$, while $i,j,k,...$ correspond to spatial indices, ranging from $1$ to $2n$. Therefore, coordinates on $\mathcal{M}=\mathbb{R}\times\Sigma$ can be split in the form $x^{\mu}=(x^{0},x^{i})$ were $x^{0}$ is the temporal coordinate associated with $\mathbb R$ and $x^{i}$ are coordinates on $\Sigma.$}
\begin{align} \label{split}
A &= A_{\mu}^{a}T_{a}dx^{\mu}=A_{0}^{a}T_{a}dx^{0}+A_{i}^{a}T_{a}dx^{i} \ .
\end{align}
The CS action \eqref{CSa} can be written in Hamiltonian form \cite{Banados:1995mq,Miskovic:2005di},
\beq\label{IH}
I_{H}\left[A\right]=\int dx^0\int_{\Sigma}d^{2n}x\;\left[\mathcal{L}_{a}^{i}\left(A\right)\dot{A}_{i}^{a}+A_{0}^{a}K_{a}\left(A\right)\right] \ ,
\eeq
where the Gauss law constraint $K_{a}$ is given by\footnote{Note that the spatial Levi-Civita symbol in $2n$ dimensions is defined in terms of the $(2n+1)$-dimensional one by $\epsilon^{i_{1}i_{2}\cdots i_{2n-1}i_{2n}}\equiv\epsilon^{0i_{1}i_{2}\cdots i_{2n-1}i_{2n}}$.}
\beq
K_{a}\left(A\right)=\frac{\kappa}{2^{n}}\left(n+1\right)\epsilon^{i_{1}i_{2}\cdots i_{2n-1}i_{2n}}g_{ab_{1}\cdots b_{n}}F_{i_{1}i_{2}}^{b_{1}}\cdots F_{i_{2n-1}i_{2n}}^{b_{n}} \ , \label{K}
\eeq
and $\mathcal{L}_{a}^{i}$ is a function of $A_i$ that determines the phase space symplectic form
\beq\label{sf}
\Omega^{ij}_{ab}=\dfrac{\delta \mathcal{L}_{b}^{j}}{\delta A_{i}^{a}} -\dfrac{\delta \mathcal{L}_{a}^{i}}{\delta A_{j}^{b}}=-\dfrac{\kappa \,n\left(n+1\right)}{2^{n-1}}\epsilon^{ijk_1k_2\cdots k_{2n-3}k_{2n-2}}g_{abc_1 \cdots c_{n-1}}F_{k_1k_2}^{c_1}\cdots F_{k_{2n-3}k_{2n-2}}^{c_{n-1}} \ .
\eeq
The phase space of the theory is defined by the spatial components of the connection $A^{a}_i$ and its canonically conjugated momenta $\pi_{a}^i$, with canonical Poisson brackets\footnote{We define canonical Poisson brackets for $\pi^0_a$ and $A_0^a$ as well, since they are canonical variables at the beginning of the analysis and variations with respect to $A_0^a$ will be useful later to generate surface integrals in the constraints of a TrFT. However, one should keep in mind that  $A_0^a$ is not a physical variable. Indeed, it is completely arbitrary due to the primary first-class constraint $\phi^0_a\approx0 $. Introducing a new constraint to gauge-fix $A_0^a$ makes $\phi^0_a$ second class, and implementing a Dirac bracket associated to that pair of constraints is equivalent to using a standard Poisson bracket only in terms of $A_i^a$ and $\pi^i_a$.}
\beq
\left\{ A_{\mu}^{a}(x),\pi_{b}^{\nu}\left(y\right)\right\} =\delta_{b}^{a}\delta_{\mu}^{\nu}\delta^{\left(2n\right)}\left(x-y\right)\,.
\eeq
When applying Dirac's Hamiltonian formalism \cite{Dirac,Henneaux-Teitelboim}, the definition of canonical momenta leads to the following primary constraints
\begin{eqnarray}
\phi^{0}_a&=&\pi^{0}_a\approx0\label{pric1} \ , \\
\phi^{i}_a&=&\pi^{i}_a-\mathcal{L}_{a}^{i}(A)\label{pric2}\approx0 \  ,
\end{eqnarray}
whereas the corresponding primary Hamiltonian reads
\beq
H = H_{0} + \int_{\Sigma}d^{2n}x\;\left[\lambda^a \phi^{0}_a+\Lambda^{a}_i\phi^{i}_a\right] \ ,
\eeq
where
\begin{equation*}
H_{0}= - \mathcal{\kappa}\int_{\Sigma}d^{2n}x\;\,A_{0}^{a}K_{a}\left(A\right) \ ,
\end{equation*}
is the canonical Hamiltonian and $\lambda^a$, $\Lambda^{a}_i$ are Lagrange multipliers. Time preservation of $\phi^{0}_a$ generates the secondary constraint $K_{a}(A)\approx0$, while time preservation of $\phi^{i}_a$ leads to no new constraints, but fixes the Lagrange multipliers $\Lambda_{i} \approx D_{i}A_{0}$ in the primary Hamiltonian. The first-class constraints of the theory can be found by redefining $K_a$ in the form
\beq\label{genG}
G_a=K_a+D_i\phi_a^{i} \; ,
\eeq
which generates the correct infinitesimal gauge transformations
\beq
\delta A^a_i(x)= \left\{ A_i^a(x), \int_\Sigma d^{2n}y\; \,\eta^b (y)G_b (y)\right\}= -D_i \eta^a (x)\,.
\eeq
Moreover, together with the second-class constraints $\phi_a^{i}$, the constraints satisfy the Poisson algebra
\beq\label{constalg}
\{G_a,G_b\}=f^c_{ab}G_c  \, ,  \quad
\{G_a,\phi_b^{i}\}=f^c_{ab}\phi_c^{i} \, , \quad
\{\phi_a^{i} , \phi_b^{j}\}=\Omega_{ab}^{ij} \,.
\eeq
The CS actions are also invariant under diffeomorphisms
\beq\label{LA}
\delta A^a_i(x) =\left\{ A_i^a(x), \int_\Sigma d^{2n}y\; \,\zeta^i (y)L_i (y)\right\} = - \mathfrak{L}_{\zeta}A^a_i \ ,
\eeq
where $\mathfrak L_\zeta$ stands for the Lie derivative along a spatial vector field $\zeta=\zeta^i \partial_i$. Using the identity
\beq\label{DevLieA}
\mathfrak{L}_{\zeta}A^a_i
=-F^{a}_{ij}\zeta^{j}+D_{i}\left(\zeta^{j}A^{a}_{j}\right) \ ,
\eeq
one can see that the generator of infinitesimal spatial diffeomorphisms can be constructed as the following combination of the constraints
\beq\label{Ldiff}
L_{i}= \mathcal{H}_{i} +A^{a}_{i}G_{a} \ , \qquad
\mathcal{H}_{i}=-F^{a}_{ij}\phi_{a}^{j} \ .
\eeq
Note that $\mathcal H_i$ generates the so-called \emph{improved spatial diffeomorphisms} \cite{Jackiw1978}, namely, $\delta A_i=-F_{ij}\zeta^j$. Time-like diffeomorphisms, on the other hand, can be shown to be generated by a combination of $G_a$ and $\mathcal{H}_i$ \cite{Banados:1995mq}.
From \cref{Ldiff} it is clear that a diffeomorphism with parameter $\zeta^i$ can be written as an improved diffeomorphism plus a gauge transformation with a field-dependent parameter $\eta^{a} = \zeta^i A^a_i$.

The dynamical structure of the $(2+1)$-dimensional theory is quite simple in comparison with higher dimensional cases. In fact, for $n=1$, the symplectic form in \cref{sf} can be inverted, implying that the constraints $\phi_{a}^{i}$ are second-class and can be set strongly to zero after implementing Dirac brackets. This leads to a canonical bracket for the reduced-phase space variables $A^{a}_i$. Once the constraints  $\phi_{a}^{i}$  are eliminated, the gauge generator \eqref{genG} reduces to the Gauss law constraint $G_a=K_a$, while $\mathcal{H}_i$ vanishes, implying that in three space-time dimensions diffeomorphisms are on-shell equivalent to gauge transformations. For $n\geqslant 2$, on the other hand, the symplectic form $\Omega_{ab}^{ij}$ is degenerate \cite{Banados:1995mq,Banados:1996yj,Miskovic:2005di} (see also \cite{Saavedra:2000wk,deMicheli:2012ye,Canfora:2014xca,Ferreira:2022ash} for details on degenerate Hamiltonian systems), which means that in higher odd-dimension, the phase space of CS theories is divided in sectors that are dynamically disconnected. Furthermore, the constraints may be irregular \cite{Miskovic:2003ex}. In generic and regular sectors, there are $2n$ null eigenvectors $(v_i)_j=F_{ij}$. This leads the first-class constraints $\mathcal H_i$ in \cref{Ldiff}, which are independent of the gauge generators $G_a$. In non-generic sectors, there are even more first-class constraints, which generate so-called accidental symmetries. These will not be considered here.

\subsection{Poisson brackets and boundary terms}
\label{section_brackets}

In order to apply Dirac formalism to a TrFT and keep contributions coming from the boundary term \eqref{Bterm} throughout, we will not use the standard Regge-Teitelboim prescription \cite{Regge:1974zd} based on regularized functionals, as it requires introducing a different boundary term in the action \eqref{action} depending on the boundary conditions chosen for the gauge fields \cite{Benguria:1976in,Henneaux:1985tv,Brown:1986nw,Brown:1986ed} (see \cite{Troessaert:2013fba} for a detailed analysis). Instead, we will consider a generalized Poisson bracket that includes the boundary variations of the relevant functionals in a consistent way. Brackets of this kind were proposed first by Soloviev \cite{Soloviev:1993zf}, and later on by Bering \cite{Bering:1998fm} in the study of Hamiltonian systems defined on spaces with boundaries.

Let us consider a functional $F$  of some set of fields $ \psi^{A} $ and their derivatives $ \partial_i\psi^{A} $, defined on a $d$-dimensional spatial section $\Sigma$ with boundary $\partial\Sigma$
\beq\label{functional}
F\left[\psi\right]=\int_{\Sigma}d^{d}x\,\mathcal{F}\left(\psi^{A}(x),\partial_{i}\psi^{A}(x)\right) +\int_{\partial \Sigma} d^{d-1}x\; \hat n_i \,f^i \left(\psi^{A}(x),\partial_{i}\psi^{A}(x)\right) \ ,
\eeq
where $\hat{n}_i$ is a unitary vector normal to $\partial\Sigma$.
This means that the variation of $F$ with respect to $ \psi^{A} $ drops a boundary term of the form
\beq
\begin{aligned}\label{variation}
\delta F\left[\psi\right]&=\int_{\Sigma}d^{d}x\left[
\frac{\partial\mathcal{F}}{\partial\psi^{A}}
-\partial_{i} \left(
\frac{\partial \mathcal{F}}{\partial \left(\partial_{i} \psi^{A}\right)}\right)
\right]\delta\psi^{A}\\[6pt]
&+\int_{\partial\Sigma}d^{d-1}x\;\hat{n}_i\,
\left[
\frac{\partial f^i}{\partial\psi^{A}}
-\partial_{j} \left(
\frac{\partial f^i}{\partial \left(\partial_{j} \psi^{A}\right)}\right)
+\frac{\partial \mathcal{F}}{\partial \left(\partial_{i} \psi^{A}\right)}
\right] \delta\psi^{A} \ .
\end{aligned}
\eeq
Usually, in this case, one would like to add a surface integral to $F$ such that, provided certain boundary conditions on the fields $ \psi^{A} $, $\delta F$ does not drop any boundary term. This would ensure the closure of Poisson brackets \cite{Brown:1986ed}.
Nevertheless, if this procedure is to be applied to the transgression action, one should dump the boundary terms \eqref{Bterm} and later add regularizing boundary terms to the first-class constraints of the theory. Eliminating this boundary term turns \cref{action} into the sum of two CS forms, yielding to the usual Hamiltonian formalism for CS theories described in Section \ref{CSanalysis}. Here, instead, we would like to exploit the properties of the transgression form by keeping its boundary term untouched and dragging its information through the Dirac formalism. Thus, we will not adopt any boundary conditions for the fields and keep all the boundary terms appearing in the variations of the functionals.
When considering variations of the form \eqref{variation}, however,  the usual Poisson bracket does not fulfill the Jacobi identity. The reason for this is that functional derivatives do not commute in the presence of a boundary \cite{Soloviev:1992be}. A Poisson bracket that satisfies the Jacobi identity and takes into account these boundary contributions can be constructed by following the works of Soloviev and Bering in references \cite{Soloviev:1993zf,Bering:1998fm}. \footnote{A similar bracket was previously proposed in \cite{LMMR} for the study of surface waves in ideal fluids.} In order to do that, the bulk and boundary parts of \cref{variation} are to be considered as the bulk and boundary functional derivatives of $F$, denoted by $\hat{\delta}F / \delta\psi^{A}$ and $\check{\delta}F / \delta\psi^{A}$, respectively,
\beq\label{variationF}
\delta F\left[\psi\right]=\int_{\Sigma}d^{d}x\;\frac{\hat{\delta}F}{\delta\psi^{A}}\delta\psi^{A}+\int_{\partial\Sigma}d^{d-1}x\;\frac{\check{\delta}F}{\delta\psi^{A}}\delta\psi^{A} \ ,
\eeq
where
\beq\label{variations}
\frac{\hat{\delta}F}{\delta\psi^{A}} =
 \frac{\delta \mathcal{F}}{\delta\psi^{A}}
 ,\qquad
\frac{\check{\delta}F}{\delta\psi^{A}}=\hat{n}_i
\left( \frac{\delta f^i}{\delta\psi^{A}}
+\frac{\partial \mathcal{F}}{\partial \left(\partial_{i} \psi^{A}\right)}
\right)  \ .
\eeq
Here we have used the notation
\beq
\frac{\delta X}{\delta \psi^A}=\frac{\partial X}{\partial\psi^{A}}
-\partial_{j} \left(
\frac{\partial X}{\partial \left(\partial_{j} \psi^{A}\right)}\right) \ .
\eeq
Then, given a symplectic structure for the phase space of the theory $\omega_{AB}$, the Poisson bracket of $F$ with some field $\psi^A$ can then be extended to the boundary, i.e.
\beq\label{extp}
\left\{ F,\psi^A (x)\right\}=\int_{\Sigma}d^{d}y\;\frac{\hat{\delta}F(y)}{\delta\psi^{B}(x)}\omega^{BA}+\int_{\partial\Sigma}d^{d-1}y\;\frac{\check{\delta}F(y)}{\delta\psi^{B}(x)}\omega^{BA} \ ,
\eeq
where $\omega^{AB}$ is the inverse of $\omega_{AB}$, and defines the Poisson bracket
\beq\label{PBpsi}
\{\psi^A(x),\psi^B(y)\}=\omega^{AB} \delta^{(d)}(x-y) \ .
\eeq
For a function $\mathcal F(x)$ depending of $\psi^A$ and its derivatives, one finds
\beq\label{PBFpsi}
\bal
&\{\mathcal F(x),\psi^A(y)\}= \int_{\Sigma} d^{d} z
\left[ \frac{\partial \mathcal F(x)}{\partial\psi^B(z)} \{\psi^B(z),\psi^A(y)\}
+
 \frac{\partial \mathcal F(x)}{\partial(\partial_i^{(z)}\psi^B(z))} \{\partial_i^{(z)}\psi^B(z),\psi^A(y)\}
\right]
\\
&= \frac{\delta \mathcal F(x)}{\delta\psi^B(y)} \omega^{BA} + \int_{\Sigma}d^{d}z\,\partial_{i}^{(z)}\left(\frac{\partial\mathcal{F}(x)}{\partial(\partial_{i}^{(z)}\psi^{B}(z))}\,\omega^{BA}\,\delta^{(D)}(z-y)\right) \ .
\eal
\eeq
Thus, for two functions $\mathcal F$ and $\mathcal G$, this leads to
\beq\label{PBFxGy}
\bal
&\{\mathcal F(x),\mathcal G(y)\}= \int_{\Sigma} d^{d}z
\frac{\delta \mathcal F(x)}{\delta\psi^B(z)} \{\psi^B(z),\mathcal G(y)\}
+\int_{\Sigma} d^{d}z \;\partial_{i}^{(z)} \left(
 \frac{\partial \mathcal F(x)}{\partial(\partial_i^{(z)}\psi^B(z))} \{\psi^B(z),\mathcal G(y)\}\right)
\\
&= \int_{\Sigma} d^{d}z\bigg[
\frac{\delta \mathcal F(x)}{\delta\psi^A(z)}
 \omega^{AB}
\frac{\delta \mathcal G(y)}{\delta\psi^B(z)}
+
\partial_{i}^{(z)}
\left(\frac{\delta \mathcal F(x)}{\delta\psi^A(z)}
\omega^{AB}
 \frac{\mathcal G(y)}{\partial(\partial_i^{(z)}\psi^B(z))}
+
 \frac{\partial \mathcal F(x)}{\partial(\partial_i^{(z)}\psi^A(z))}  \omega^{AB}
\frac{\delta \mathcal G(y)}{\delta\psi^B(z)}
\right) \bigg] \ .
\eal
\eeq
On the other hand, for two functionals of the form \eqref{functional}, i.e.
\beq
\bal
F\left[\psi\right]&=\int_{\Sigma}d^{d}x\,\mathcal{F}(x)
+\int_{\partial \Sigma} d^{d-1}x\;\hat n_i\,f^i(x) \ ,
\\
G\left[\psi\right]&=\int_{\Sigma}d^{d}y\,\mathcal{G}(y)
+\int_{\partial \Sigma} d^{d-1}y\;\hat n_i\,g^i(y) \ ,
\eal
\eeq
one finds
\beq\label{PBFG}
\bal
&\{F,G\}=\int_{\Sigma} d^{d}x \int_{\Sigma} d^{d}y \{\mathcal F(x),\mathcal G(y)\}
+\int_{\Sigma} d^{d}x \int_{\partial\Sigma} d^{d-1}y \;\hat n_i\, \left(\{\mathcal F(x), g^i (y)\}
+\{ f^i (y), \mathcal G (x)\} \right)
\\
&=\int_{\Sigma} d^{d}x \int_{\Sigma} d^{d}y \{\mathcal F(x),\mathcal G(y)\}
+\int_{\Sigma} d^{d}x \int_{\partial\Sigma} d^{d-1}y \;\hat n_i\, \left(
\frac{\delta \mathcal F(x)}{\delta\psi^A(z)}
 \omega^{AB}
\frac{\delta g^i(y)}{\delta\psi^B(z)}
+\frac{\delta f^{i}(x)}{\delta\psi^A(z)}
 \omega^{AB}
\frac{\delta G(y)}{\delta\psi^B(z)} \right) \ .
\eal
\eeq
Note that in the second line, we have used the fact that  $\partial\partial\Sigma=\emptyset$. Replacing \eqref{PBFxGy} in the bulk integral of \cref{PBFG} and using the definitions \eqref{variations} we find
\beq
\bal
\{F,G\} &= \int_{\Sigma} d^{d}x \frac{\hat\delta F}{\delta\psi^A(x)}
 \omega^{AB}
\frac{\hat \delta G}{\delta\psi^B(x)}
\\
&+\int_{\partial\Sigma} d^{d-1}x \left( \frac{\hat\delta F}{\delta\psi^A(x)}
 \omega^{AB}
\frac{\check \delta G}{\delta\psi^B(x)}
 +
\frac{\check\delta F}{\delta\psi^A(x)}
 \omega^{AB}
\frac{\hat \delta G}{\delta\psi^B(x)}
\right) \ .
\eal
\eeq

Considering a set of fields and their corresponding momenta, $\psi^A=\left(\theta^{\alpha}, \pi_{\alpha} \right)$, with canonical (equal time) Poisson brackets given by
\beq\label{pb}
\begin{array}{lcl}
\left\{ \theta^{\alpha}(x),\pi_{\beta}(y)\right\}&=&\delta^{(d)}(x-y) \ ,
\end{array}
\eeq
the definition \eqref{extp} implies that the extended Poisson bracket of two functions $F$ and $G$ of the form \eqref{functional} is given by
\begin{align}\label{soloviev}
\left\{ F,G\right\}  & =\int_{\Sigma}d^{d}x\left(\frac{\hat{\delta}F}{\delta\theta_{\alpha}}\frac{\hat{\delta}G}{\delta\pi^{\alpha}}-\frac{\hat{\delta}G}{\delta\theta_{\alpha}}\frac{\hat{\delta}F}{\delta\pi^{\alpha}}\right)\nonumber \\
 & +\int_{\partial\Sigma}d^{d-1}x\left(\frac{\check{\delta}F}{\delta\theta_{\alpha}}\frac{\hat{\delta}G}{\delta\pi^{\alpha}}+\frac{\hat{\delta}F}{\delta\theta_{\alpha}}\frac{\check{\delta}G}{\delta\pi^{\alpha}}\right)-\int_{\partial\Sigma}d^{d-1}x\left(\frac{\check{\delta}G}{\delta\theta_{\alpha}}\frac{\hat{\delta}F}{\delta\pi^{\alpha}}+\frac{\hat{\delta}G}{\delta\theta_{\alpha}}\frac{\check{\delta}F}{\delta\pi^{\alpha}}\right) \ .
\end{align}
Notice that the bracket \eqref{soloviev} evaluated on functions with only bulk variations reduces to the standard Poisson bracket. We should also stress here that, in the case of a first-order theory, such as a TrFT, the bracket \eqref{soloviev} is obtained by keeping all the boundary terms when integrating by parts and applying the standard Poisson bracket inside those boundary terms. This is another way to understand the rather formal formula for the generalized bracket given in \eqref{soloviev}.

\subsection{Dirac algorithm for transgression field theory}

Now we turn back to the transgression action \eqref{action}. As stated in \Cref{transgression}, we will consider the case where $\mathcal{M}=\bar{\mathcal{M}}$, but the results can be straightforwardly generalized to the case of cobordant manifolds. Following the same steps as in the Hamiltonian construction of CS theories reviewed in \Cref{CSanalysis}, we use the splitting \eqref{split} to write
\beq\label{split2}
\bal
A &=A_\mu^a T_a dx^\mu = A_0^a T_a dx^0 + A_i^a T_a dx^i \ ,
\\[4pt]
\bar A & =\bar A_\mu^a T_a dx^\mu=\bar A_0^a T_a dx^0+\bar A_i^a T_a dx^i \ ,
\eal
\eeq
and re-derive the analog of \cref{IH}, this time keeping all the surface integrals. In the presence of a boundary, the $(2n+1)$-dimensional CS action \eqref{CSa} can be written as the bulk term \eqref{IH} plus a surface integral, namely
\beq\label{csah}
\kappa \int_{\mathcal M}CS_{2n+1}\left[A\right]=I_{H}[A]-\int dx^0\int_{\partial\Sigma}ds_{i}\,\mathcal{L}_{a}^{i}\left(A\right)A_{0}^{a} \ .
\eeq
Notice that, until now, we have dropped the boundary term in the CS action \eqref{csah} and studied only \cref{IH}. In the Regge-Teiteilboim formalism, this boundary term is generally neglected and, provided some boundary conditions, a different boundary term must to be added to the action in order to render its variation well-defined. In the following, instead, we will study the transgression action by keeping all the boundary terms appearing throughout Dirac formalism.

By splitting the spatial and temporal components of $A$ and $\bar A$ according to \cref{split2}, the boundary term $\mathcal{B}_{2n}$ defined in \cref{Bterm} takes the following general form\footnote{To avoid cluttering equations, we have introduced the notation $f(A,\bar A)- \{A\leftrightarrow\bar A\}\equiv f(A,\bar A)-f(\bar A, A)$. Written at the end of an equation, as in \cref{bt}, the symbol $\{A\leftrightarrow\bar A\}$ refers to all the terms on the right-hand side written before it.}
\beq\label{bt}
\int_{\partial\mathcal{M}}\mathcal{B}_{2n}\left[A,\bar{A}\right]
= \int dx^0 \int_{\partial\Sigma} \, ds_{i} \, \Big[\mathcal{R}_{a}^{i}(A,\bar{A})A_{0}^{a}
+ \rho_{a}^{ij}(A,\bar{A})\dot{A}_{j}^{a}\Big] \; - \; \Big\{A\leftrightarrow\bar{A}\Big\} \ ,
\eeq
where $\mathcal{R}_{a}^{i}$ and $\rho_{a}^{ij}$ are functions of the gauge connections. Therefore, the $(2n+1)$-dimensional transgression action \eqref{action} looks like
\beq\label{actionh}
I\left[A, \bar A\right] = I_{H}[A] - \int dx^0 \, \int_{\partial\Sigma} d^{2n-1}x\,\hat{n}_{i} \left[\left(\mathcal{L}_{a}^{i}(A) + \mathcal{R}_{a}^{i}(A,\bar{A})\right)A_{0}^{a}+\rho_{a}^{ij}(A,\bar{A})\dot{A}_{j}^{a}\right] \; - \; \Big\{A\leftrightarrow\bar{A}\Big\} \ ,
\eeq
where we have defined the boundary surface element as $ds_{i}=d^{2n-1}x\,\hat{n}_{i}$, and $\hat{n}_{i}$ is the unit vector normal to $\partial \Sigma$. The non-vanishing canonical Poisson brackets (at equal time) for the canonical variables are given by
\beq\label{canonical}
\left\{A_{\mu}^{a}(x),\pi_{b}^{\nu}(y)\right\} = \delta_{b}^{a}\delta_{\mu}^{\nu}\delta^{(2n)}(x-y) \ , \quad
\left\{ \bar{A}_{\mu}^{a}(x),\bar{\pi}_{b}^{\nu}\left(y\right)\right\} =\delta_{b}^{a}\delta_{\mu}^{\nu}\delta^{\left(2n\right)}\left(x-y\right) \ .
\eeq

The smeared primary constraints, obtained from the canonical momenta associated with the action in \cref{actionh}, are given by
\beq\label{primary}
\begin{aligned}
\phi\left[\eta\right]&=\int_\Sigma d^{2n}x\;\eta^{a}\pi_{a}^{0} \ , \\
\bar{\phi}\left[\bar{\eta}\right]&=\int_\Sigma d^{2n}x\;\bar{\eta}^{a}\bar{\pi}_{a}^{0} \ , \\
\Phi[\Upsilon]&=\int_{\Sigma} d^{2n}x \, \Upsilon_{i}^{a}\left(\pi_{a}^{i}-\mathcal{L}_{a}^{i}(A)\right) + \int_{\partial\Sigma} d^{2n-1}x\,\hat{n}_{i} \, \Upsilon_{j}^{a} \, \rho_{a}^{ij}(A,\bar{A}) \  , \\
\bar{\Phi}[\bar{\Upsilon}] & = \int_{\Sigma} d^{2n}x\;\bar{\Upsilon}_{i}^{a}\left(\bar{\pi}_{a}^{i} + \mathcal{L}_{a}^{i}(\bar{A})\right)-\int_{\partial\text{\ensuremath{\Sigma}}}d^{2n-1}x\,\hat{n}_{i}\,\bar{\Upsilon}_{j}^{a}\rho_{a}^{ij}\left(\bar{A},A\right) \ ,
\end{aligned}
\eeq
while the canonical Hamiltonian reads\footnote{The Hamiltonian associated with a transgression form has been previously studied in \cite{Aviles:2016raj}, where it was shown that it satisfies a triangular equation analogous to \eqref{transgression_definition_2}.}
\beq\label{canonical_Hamiltonian}
\begin{aligned}
H_{0}  = - \int_{\Sigma}d^{2n}x\;K_{a}(A)A_{0}^{a}
 + \int_{\partial\Sigma}d^{2n-1}x\,\hat{n}_{i}\left[\mathcal{L}_{a}^{i}\left(A\right)+\mathcal{R}_{a}^{i}\left(A,\bar{A}\right)\right] A_{0}^{a}\;-\;\Big\{A\leftrightarrow\bar{A}\Big\} \ .
\end{aligned}
\eeq
The primary Hamiltonian is obtained by adding a linear combination of the constraints to \cref{canonical_Hamiltonian}, and reads
\beq\label{H-1}
H = H_{0} + \phi[\lambda] + \bar{\phi}[\bar{\lambda}] + \Phi[\Lambda] + \bar{\Phi}[\bar{\Lambda}] \ ,
\eeq
where $\lambda^{a}$, $\bar{\lambda}^{a}$, $\Lambda_{i}^{a}$, $\bar{\Lambda}_{i}^{a}$ are Lagrange multipliers. The bulk and boundary variations of the primary constrains can be computed using the definitions \eqref{variationF}, where now we use the collective notation $\psi^A = \left( A^a_i, \bar A^a_i , \pi_a^i , \bar \pi_a^i \right)$.
This leads to the following non-vanishing Poisson brackets
\beq
\begin{aligned}
 \{\Phi[\eta],\Phi[\Lambda]\} &= \int_\Sigma d^{2n}x\; \eta^a_i\Lambda^b_j\Omega_{ab}^{ij}(A) + \int_{\partial\Sigma} d^{2n-1}x\,\hat{n}_{k}\,\eta^a_i\Lambda^b_j \omega_{ab}^{ijk}(A,\bar{A})   \ , \\
 \{\bar{\Phi}[\bar{\eta}],\bar{\Phi}[\bar{\Lambda}]\} &= - \int_\Sigma d^{2n}x\; \bar{\eta}^a_i\bar{\Lambda}^b_j\Omega_{ab}^{ij}(\bar{A}) - \int_{\partial\Sigma} d^{2n-1}x\,\hat{n}_{k}\, \bar{\eta}^a_i\bar{\Lambda}^b_j \omega_{ab}^{ijk}(\bar{A},A)  \ , \\
 \{\Phi[\eta],\bar{\Phi}[\bar{\Lambda}]\} &= \int_{\partial\Sigma} d^{2n-1}x\,\hat{n}_{i}\, \eta^a_j \bar{\Lambda}^c_k \zeta^{ijk}_{ba} \ ,
 \end{aligned}
 \eeq
 where
 \beq
\begin{aligned}
 &\omega^{kij}_{ab}(A,\bar{A}) = -\frac{\check\delta\mathcal{\rho}_{b}^{kj}(A,\bar{A})}{\delta A_{i}^{a}} + \frac{\check\delta\mathcal{\rho}_{a}^{ki}(A,\bar{A})}{\delta A_{j}^{b}}-\frac{\partial\mathcal{L}_{a}^{i}(A)}{\partial\left(\partial_{k}A_{j}^{b}\right)} \ , \\
& \zeta^{ijk}_{ba}=\frac{\check\delta \rho^{ij}_a(A,\bar{A})}{\delta \bar{A}^b_k} + \frac{\check\delta \rho^{ik}_b(\bar{A},A)}{\delta A^a_j} \ ,
\end{aligned}
\eeq
and $\Omega^{ij}_{ab}$ is the symplectic form defined in \cref{sf}.

Time preservation of the primary constraints $\phi\lbrack\eta]$ and $\bar{\phi}[\bar{\eta}]$ leads to secondary constraints, given by
\beq\label{secondclass}
\begin{aligned}
\chi\left[\eta\right]&=\left\{ \phi[\eta],H\right\} =\int d^{2n}x\;\eta^{a}K_{a}\left(A\right)-\kappa\int_{\partial\Sigma} d^{2n-1}x\,\hat{n}_{i}\,\eta^{a}\left(\mathcal{L}_{a}^{i}\left(A\right)+\mathcal{R}_{a}^{i}\left(A,\bar{A}\right)\right) \ ,\\
\bar{\chi}\left[\bar{\eta}\right]&=\left\{ \bar{\phi}[\bar{\eta}],H\right\} =-\int d^{2n}x\;\bar{\eta}^{a}K_{a}\left(\bar{A}\right)+\mathcal{\kappa}\int_{\partial\Sigma}d^{2n-1}x\,\hat{n}_{i}\,\bar{\eta}^{a}\left(\mathcal{L}_{a}^{i}\left(\bar{A}\right)+\mathcal{R}_{a}^{i}\left(\bar{A},A\right)\right) \ .
\end{aligned}
\eeq
In general, even though the set of constraints $\left\{ \chi\lbrack\eta],\Phi\lbrack\Upsilon],\bar{\chi}[\bar{\eta}],\bar{\Phi}[\bar{\Upsilon}]\right\} $
do not have vanishing Poisson brackets among themselves, they have a degenerate Poisson bracket matrix. This implies that they do not form a second-class set, but there are certain combinations of them that are first-class. In fact, the following combination of the constraints
\beq\label{gamma}
\Gamma\left[\eta\right]=\chi[\eta]-\Phi[D\eta] \ , \qquad
\bar{\Gamma}\left[\bar{\eta}\right]=\bar{\chi}[\bar{\eta}]-\bar{\Phi}[\bar{D}\bar{\eta}] \ ,
\eeq
can be shown to be first-class \cite{Banados:1995mq,Miskovic:2005di}.\footnote{The notion of first- and second-class constraints here refers only to the bulk dynamics, i.e. up to boundary terms.} It is important to remark that, similarly to what happens in higher dimensional CS theory, for dimensions greater than three, the transgression theory has a degenerate symplectic form, implying that there are regions in phase space, where there might be even more first-class constrains due to accidental symmetries \cite{Banados:1995mq,Banados:1996yj}. As said above, these extra first-class constraints will not be considered here.

Notice that, in these expressions, $A_{0}$ and $\bar{A}_{0}$ play the role of gauge parameters. Thus, the condition \eqref{gauge_transf_infinitesimal} necessary for gauge invariance leads to
\beq\label{condA0}
A_{0}\bigg|_{\partial\Sigma}=\bar{A}_{0}\bigg|_{\partial\Sigma}\,.
\eeq
The total Hamiltonian \eqref{H-1} can be written
in terms of \cref{gamma} as
\beq\label{H3}
H=\Gamma\left[A_{0}\right]+\bar{\Gamma}\left[\bar{A}_{0}\right]+\Phi[\Lambda-D A_0]+\bar \Phi[\bar \Lambda-\bar D \bar A_0] \,.
\eeq
As we will see in the following section, the term $\Gamma\left[A_{0}\right]+\bar{\Gamma}\left[\bar{A}_{0}\right]$ corresponds to the gauge symmetry generator of the theory provided the condition \eqref{condA0} holds. Since the constraints $\Phi$ and $\bar\Phi$ transform in the coadjoint representation of the gauge group \cite{Banados:1995mq,Miskovic:2005di}, the constraints \eqref{gamma} are automatically preserved during the time evolution of the system, and thus there are no more constraints in the theory.

As discussed in \Cref{CSanalysis}, for $D=2+1$ the theory is generic. Temporal preservation of the constraints \eqref{gamma} leads to the conditions $\Lambda_{i}  \approx  D_{i}A_{0}$ and $\bar{\Lambda}_{i}  \approx  \bar{D}_{i}\bar{A}_{0}$. In this exceptional case, the constraints $\Phi[\Upsilon]$ and $\bar{\Phi}[\bar{\Upsilon}]$ are second-class and do not acquire boundary contributions (see Section \ref{section_2+1}). The first-class constraints of the theory are $\chi[\eta]$ and $\bar \chi[\bar \eta]$, which are equivalent to \eqref{gamma} after implementing the Dirac bracket that allows one to eliminate the second-class constraints. For $D\geq4+1$, however, the degeneracies of the symplectic form imply that only some of the Lagrange multipliers are fixed. Then, as mentioned above, the existence of arbitrary parameters in the total Hamiltonian leads to more first-class generators, associated with accidental symmetries that appear in the degenerate sectors of the theory.

\section{Symmetry generators and conserved charges}
\label{3y5D}

Now, let us consider the following combination of the first-class constraints defined in \cref{gamma},
\beq\label{g_functional}
\mathcal{G}\left[\eta,\bar{\eta}\right]=\Gamma\left[\eta\right]+\bar{\Gamma}\left[\bar{\eta}\right]{\rm =bulk}+\mathcal{Q}\left[\eta,\bar{\eta}\right] \ ,
\eeq
where the bulk part weakly vanishes, and the surface integral is given by
\beq\label{Q}
\begin{aligned}
\mathcal{Q}\left[\eta,\bar{\eta}\right] &=  -\kappa\int_{\partial\Sigma}d^{2n-1}x\,\hat{n}_{i}\Bigg[  \eta^a\bigg(\mathcal{L}^i_a(A)+\mathcal{R}^i_a(A,\bar{A})-D_j\rho^{ij}_a(A,\bar{A}) \bigg) \\
&\hskip3.2truecm - \bar{\eta}^a\bigg( \mathcal{L}^i_a(\bar{A})+\mathcal{R}^i_a(\bar{A},A)-\bar{D}_j\rho^{ij}_a(\bar{A},A) \bigg) \Bigg] \ .
\end{aligned}
\eeq
From \cref{gauge_transf_infinitesimal}, we know that the gauge invariance of the action is guaranteed even when the connections $A$ and $\bar{A}$ transform with different gauge parameters, provided they are identified at the boundary. Therefore, the generator of gauge transformations is given by
\beq\label{gauge_functional}
G\left[\eta,\bar{\eta}\right]  =  \mathcal{G}\left[\eta,\bar{\eta}\right]
\bigg\vert _{\left.\eta\right\vert _{\partial\Sigma}=\left.\bar{\eta}\right\vert _{\partial\Sigma}} \ .
\eeq
Furthermore, the generator \eqref{gauge_functional} clearly gives the correct gauge transformation for the gauge connections \eqref{gauge_transf_infinitesimal} when using the generalized bracket \eqref{soloviev},
\beq
\begin{aligned}
\delta A^a_i  &=\left\{ A^a_i,G\left[\eta,\bar\eta\right]\right\} = -D_i\eta^a \ ,
\\[5pt]
\delta\bar A^a_i & =\left\{ \bar A^a_i,G\left[\eta,\bar\eta\right]\right\} = -\bar D_i\bar\eta^a  \ ,
\qquad
\left.\eta^a\right\vert _{\partial\Sigma}=\left.\bar\eta^a\right\vert _{\partial\Sigma} \ .
\end{aligned}
\eeq

We mentioned in \Cref{CSanalysis} that, in the case of higher-dimensional CS theory, when restricted to a sector in which the symplectic form has maximal rank, its zero modes lead to an independent generator for spatial diffeomorphisms given by \cref{Ldiff}. In the case of a transgression field theory, the analogous smeared constraint is given by
\beq\label{diffeo_functional}
L\left[\zeta,\bar\zeta\right] = \Big(\Phi\left[\zeta \cdot F\right]+\bar{\Phi}\left[\bar \zeta \cdot \bar{F}\right]+\mathcal{G}\left[\zeta \cdot A,\bar{\zeta}\cdot \bar{A} \right]\Big)\Bigg\vert _{\left.\zeta \right\vert _{\partial\Sigma}=\left.\bar{\zeta}\right\vert _{\partial\Sigma}} \ ,
\eeq
where we use the notation $\zeta\cdot A\equiv \zeta^i A_{i}$ and $\zeta\cdot F\equiv \zeta^j F_{ij}$. This constraint generates the transformations
\beq
\begin{aligned}
\delta A_{i}^{a}(x) & =\left\{ A_{i}^{a}(x),L\left[\zeta,\bar\zeta\right]\right\} =-\mathfrak L_{\zeta}A_{i}^a \ ,
\\[5pt]
\delta\bar{A}_{i}^{a}(x) & =\left\{ \bar{A}_{i}^{a}(x),L\left[\zeta,\bar\zeta\right]\right\} =-\mathfrak L_{\bar\zeta}\bar{A}_{i}^a \ ,
\qquad
\left.\zeta^i \right\vert _{\partial\Sigma}=\left.\bar\zeta^i\right\vert _{\partial\Sigma} \ .
\end{aligned}
\eeq

By using \cref{g_functional,Q}, we can evaluate the boundary integrals of the generators given in \cref{gauge_functional,diffeo_functional}, which correspond to the surface charges associated with gauge transformations and spatial diffeomorphisms, respectively;
\beq
\bal
&Q_{\rm gauge}\left[\eta\right]  =  \mathcal{Q}\left[\eta,\bar{\eta}\right]
\bigg\vert _{\eta =\bar\eta} \ ,\\
&Q_{\rm diff}\left[\zeta\right]  = \mathcal{Q}\left[\zeta \cdot A,\bar{\zeta}\cdot \bar{A} \right]\Bigg\vert _{\zeta =\bar\zeta}
+ \int_{\partial\Sigma} d^{2n-1}x\,\hat{n}_{i} \,\zeta^k\left(  \rho_{a}^{ij}(A,\bar{A})\,F^a_{jk}
- \rho_{a}^{ij}(\bar A,A)\, \bar F^a_{jk}\right) \ .
\eal
\eeq
It is important to note that, from the general theory of conserved charges in gauge theories \cite{Barnich:2001jy} (see also \cite{Abbott:1982jh}), equivalence classes of conserved forms are known to be in one-to-one correspondence with reducibility parameters that leave the gauge fields invariant asymptotically. In the case of gauge transformations the reducibility parameters are defined by
\beq
D\eta^a\big|_{\partial\Sigma}=0=\bar D\eta^a\big|_{\partial\Sigma} \ ,
\eeq
whereas for diffeomorphisms the analog condition reads
\beq
\mathfrak L_\zeta A^a_i\big|_{\partial\Sigma}=0= \mathfrak L_\zeta \bar A^a_i\big|_{\partial\Sigma} \ .
\eeq
Thus, integrating by parts of the covariant derivatives in \cref{Q} and using the reducibility condition on $\eta^a$, $\bar \eta^a$, $\zeta^i$ and $\bar \zeta^i$ allows one to eliminate the functions $\rho^{ij}_a$ in the resulting surface integrals. This leads to
\begin{align}
&Q_{\rm gauge }\left[\eta\right]=-
\kappa\int_{\partial\Sigma}
d^{2n-1}x\,\hat{n}_{i} \,\eta^{a}\bigg[\mathcal{L}_{a}^{i}\left(A\right)+\mathcal{R}_{a}^{i}\left(A,\bar{A}\right)\;-\;\Big\{A\leftrightarrow\bar{A}\Big\}\bigg] \ ,
\label{charges_gauge_generic}
\\[5pt]
&Q_{\rm diff }\left[\zeta\right] = -
\kappa\int_{\partial\Sigma}
d^{2n-1}x\,\hat{n}_{i} \,\zeta^{j} \bigg[A_{j}^{a}\Big(\mathcal{L}_{a}^{i}\left(A\right)+\mathcal{R}_{a}^{i}\left(A,\bar{A}\right)\Big)\;-\;\Big\{A\leftrightarrow\bar{A}\Big\}\bigg] \ .
\label{charges_diffeo_generic}
\end{align}
Remarkably, these surface charges are obtained directly from the transgression form by dragging all the boundary terms in the action through Dirac formalism without the need to add regularizing boundary terms at any stage of the procedure. In the following, we show how to apply this result for a TrFT in $(2+1)$ and $(4+1)$ dimensions. The generators of gauge transformations and spatial diffeomorphisms will be constructed using the general results of \cref{gauge_functional,diffeo_functional}. In each case, the symmetry generators satisfy the Poisson algebra\footnote{Notice that on the right-hand side of \eqref{algebra_generators}, $[\![\eta,\lambda]\!]$ denotes the commutator \eqref{LieAlg} of $\mathfrak g$-valued gauge parameters, whereas $\{\zeta,\xi\}_{\rm Lie}$ stands for the Lie bracket of vector fields. The same definitions hold for the barred sector.}
\beq\label{algebra_generators}
\begin{aligned}
&\{G[\eta,\bar{\eta}],G[\lambda,\bar{\lambda}]\}=G\left[[\![\eta,\lambda]\!],[\![\bar{\eta},\bar{\lambda}]\!]\right] \ , \\[5pt]
&\left\{ L\left[\zeta,\bar\zeta\right],L\left[\xi,\bar\xi\right]\right\} = - L\left[\{\zeta,\xi\}_{\rm Lie},\{\bar\zeta,\bar\xi\}_{\rm Lie}\right] \ ,
\end{aligned}
\eeq
which can be shown using the generalized Poisson bracket \eqref{soloviev}. From \cref{algebra_generators} it follows that, after setting the constraints strongly to zero, the corresponding surface charges satisfy the Dirac algebra\footnote{Here $\{ X,Y \}^{\ast}$ denotes the Dirac bracket of $X$ and $Y$ \cite{Dirac}.}
\begin{align}
&\left\{ Q_{\rm gauge}\left[\eta\right],Q_{\rm gauge}\left[\lambda\right]\right\}^{\ast} =Q_{\rm gauge}\left[ [\![\eta,\lambda]\!]\right] \ ,
\label{algebra_gauge}
\\[5pt]
&\left\{ Q_{\rm diff}\left[\zeta\right],Q_{\rm diff}\left[\xi\right]\right\}^{\ast} =-Q_{\rm diff}\left[\{\zeta,\xi\}_{\rm Lie}\right] \ .
\label{algebra_diffeo}
\end{align}
The algebra of the charges does not include central terms, which is to be expected from the true gauge invariance of the theory. This is in contrast with CS theory, where quasi-invariance of the action under gauge transformations translates into central charges in the surface charge algebra \cite{Banados:1994tn,Banados:1996yj,Miskovic:2005di}.
One of the advantages of the expressions \eqref{charges_gauge_generic} and \eqref{charges_diffeo_generic} is that they give formulas for the conserved charges which can be directly evaluated once the functions $\mathcal L^i_a$ and $\mathcal R^i_a$ are read off from the transgression action. Furthermore, the conserved charges associated with CS theory can be obtained by setting one of the connections to be zero.

\subsection{$2+1$ dimensions}
\label{section_2+1}

As a first example, let us consider the $(2+1)$-dimensional TrFT, which is obtained by evaluating \eqref{action2} for $n=1$. This yields to
\beq\label{t2+1}
I[A,\bar{A}] = \kappa\int_{\mathcal M} \mbox{CS}_{2+1}[A] - \kappa\int_{\mathcal M} \mbox{CS}_{2+1}[\bar{A}] - \kappa\int_{\partial \mathcal{M}}\mathcal{B}_{2}[A,\bar{A}] \ ,
\eeq
where the CS terms and the boundary term follow from \cref{CSa} and \cref{Bterm}, and read
\beq \label{CS_2+1}
CS_{2+1}\left[A\right]=\left\langle dA+\dfrac{2}{3}A^3 \right\rangle \ ,\qquad
\mathcal{B}_2\left[A,\bar{A}\right]=\left\langle A\bar{A}\right\rangle \ .
\eeq
Splitting the connection $A$ as in \cref{split2} we can write these terms in the form \eqref{csah} and \eqref{bt}, from which we can directly read off the functions $\mathcal{L}_{a}^{i}$ and $\mathcal{R}_{a}^{i}$,
\beq\label{L_and_R_2+1}
\mathcal{L}_{a}^{i}(A) = \kappa\epsilon^{ij}g_{ab}A_{j}^{b} \ , \qquad \mathcal{R}_{a}^{i}(A,\bar A) = -\kappa\epsilon^{ij}g_{ab}\bar{A}_{j}^{b} \ .
\eeq
These expressions can now be plugged in \cref{charges_gauge_generic,charges_diffeo_generic} to obtain the conserved charges
\beq\label{charges_2+1}
\begin{aligned}
&Q_{\rm gauge}[\eta] = -2\kappa\int_{\partial\Sigma}dx\,\hat{n}_{i} \,\epsilon^{ij}\,g_{ab}\,\eta^{a}\,(A_{j}^{b}-\bar{A}_{j}^{b}) \ , \\
&Q_{\rm diff}[\zeta] = - \kappa\int_{\partial\Sigma}dx\,\hat{n}_{i} \,\epsilon^{ij}\,g_{ab}\,\zeta^{k}\,(A_{k}^{a}+\bar{A}_{k}^{a})\,(A_{j}^{b}-\bar{A}_{j}^{b}) \ .
\end{aligned}
\eeq
These charges are in agreement with the ones found in reference \cite{Mora:2006ka}. The $(2+1)$-dimensional case is special due to the non-degenerate structure of the symplectic form. Due to this reason, the constraints associated with these charges are not independent \cite{Banados:1994tn}. Another way to see this is that in $2+1$ dimensions the Lie derivative of a gauge connection is on-shell equivalent to a gauge transformation with a field-dependent parameter. Indeed, from \cref{DevLieA}, one finds that
\beq\label{LA2+1}
\mathfrak L_{\zeta}A^{a}_{i}
\underset{\rm on-shell}{=}D_{i}\left(\zeta^{j}A^{a}_{j}\right)
 \,.
\eeq
The generators associated with the charges \eqref{charges_2+1} can also be constructed directly from the results we have already obtained. Using the general expressions \eqref{csah} and \eqref{bt}, we see that in this case \eqref{CS_2+1} leads to
\beq
K_{a}\left(A\right) =  \kappa\epsilon^{ij}g_{ab}F_{ij}^{b} \ , \qquad
\rho_{a}^{ij} \left(A,\bar A\right) = 0\,.
\eeq
Replacing this together with \cref{L_and_R_2+1} in \cref{gauge_functional,diffeo_functional}, and using the results in \cref{primary,secondclass,gamma,g_functional}, yields
\beq\label{Ggen2+1}
\bal
G\left[\eta,\bar{\eta}\right] &= \int_{\Sigma}d^{2}x\bigg[
-D_{i}\eta^{a}\pi_{a}^{i}-\bar{D}_{i}\bar{\eta}^{a}\bar{\pi}_{a}^{i}
\\
&\hskip1.5truecm
+\kappa\mathcal{\epsilon}^{ij}g_{ab}\Big(\eta^{a}F_{ij}^{b}-\bar{\eta}^{a}\bar{F}_{ij}^{b}+D_{i}\eta^{a}A_{j}^{b}-\bar{D}_{i}\bar{\eta}^{a}\bar{A}_{j}^{b}\Big)
\bigg]
+Q_{\rm gauge}\left[\eta\right] \ ,\\[5pt]
L\left[\zeta,\bar{\zeta}\right] &= \int_{\Sigma}d^{2}x\bigg[-\mathfrak L_{\zeta}A^{a}_{i}\pi_{a}^{i}-\mathfrak L_{\bar\zeta}\bar A^{a}_{i}\bar{\pi}_{a}^{i}\\
&\hskip1.5truecm+\mathcal{\kappa}\mathcal{\epsilon}^{ij}g_{ab}\Big(\zeta^k A_k^{a}F_{ij}^{b}-\bar\zeta^k \bar A_k^{a}\bar{F}_{ij}^{b}+\mathfrak L_{\zeta}A^{a}_{i}A_{j}^{b}-\mathfrak L_{\bar\zeta}\bar A^{a}_{i}\bar{A}_{j}^{b}\Big)\bigg]
+Q_{\rm diff}\left[\zeta\right] \ .
\eal
\eeq
Finally, from the generalized bracket \eqref{soloviev}, one can explicitly show that these generators satisfy the Poisson algebra \eqref{algebra_generators}.
This automatically leads to the Dirac algebra \eqref{algebra_gauge}, \eqref{algebra_diffeo}, after setting the constraints strongly to zero. For $\bar{A}_i^a=0$, \cref{Ggen2+1} reduce to the regularized smeared generators for gauge transformations and spatial diffeomorphisms for CS theory, and the associated charges \eqref{charges_2+1} reduce to the known results that are derived using the Regge-Teitelboim method \cite{Banados:1994tn,Park:1998yw,Troessaert:2013fba}.

\subsection{$4+1$ dimensions}
\label{section_4+1}

As a second relevant case, we construct the $(4+1)$-dimensional TrFT action by evaluating \cref{action2} for $n=2$. This gives
\beq\label{t4+1}
I[A,\bar{A}] = \kappa
\int_{\mathcal M} CS_{4+1}(A)  -\kappa \int_{\mathcal M} CS_{4+1}(\bar{A})  -  \kappa\int_{\partial \mathcal{M}}\mathcal{B}_{4} \ .
\eeq
The explicit form of the CS terms follow from \cref{CSa}, from which we find
\beq\label{cs4+1}
CS_{4+1}\left[A\right] = \left\langle AdAdA+\frac{3}{2}A^{3}dA+\frac{3}{5}A^{5}\right\rangle \ ,
\eeq
whereas the boundary term \eqref{Bterm} in this case reads
\beq\label{b4+1}
\mathcal{B}_{4}\left(A,\bar{A}\right) = \left\langle A\bar{A}\left(F+\bar{F}\right)+\frac{1}{2}\bar{A}A^{3}+\frac{1}{2}\bar{A}^{3}A+\frac{1}{4}A\bar{A}A\bar{A}\right\rangle \ .
\eeq
Splitting the connections $A$ and $\bar A$ in spatial and temporal components according to \cref{split2}, these expressions can be put in the form \eqref{csah} and \eqref{bt}, where
\beq\label{LandR5D}
\bal
&\mathcal{L}_{a}^{i}\left(A\right) = \kappa \epsilon^{ijkl}g_{abc}\left( F_{jk}^bA_{l}^c-\frac{1}{4}f^b_{de}A_{j}^c A_{k}^d A_{l}^e\right) \ ,
\\[5pt]
&\mathcal{R}_{a}^{i}\left(A,\bar{A}\right) =  \mathcal{\kappa}\epsilon^{ijkl}g_{abc}\Bigg[ \frac{1}{2}A^b_{j}\bar{F}^c_{kl}-A^b_{j}f^c_{de}\left(\bar{A}^d_{k}\bar{A}^e_{l}-\frac{1}{2}\bar{A}^d_{k}A^e_{l}-\frac{1}{2}A^d_{k}\bar{A}^e_{l}\right)
\\[5pt]
 &\hskip2.3truecm-\frac{1}{2}\bar{A}_{j}\left(2F_{kl}+\bar{F}_{kl}\right)+\bar{A}_{i}\left(\frac{1}{2}A_{j}A_{k}+\frac{1}{2}\bar{A}_{j}\bar{A}_{k}-\frac{1}{2}A_{j}\bar{A}_{k}\right)\Bigg] \ .
\eal
\eeq
Using these expressions, we can write down the conserved charges associated with gauge transformations and spatial diffeomorphisms in a straightforward way from the general formulae \eqref{charges_gauge_generic} and \eqref{charges_diffeo_generic},
\beq\label{charges_4+1}
\bal
&Q_{\rm gauge}\left[\eta\right] = - \frac{\kappa}{2}\int_{\partial\Sigma}d^3x\,\hat{n}_{i} \,\epsilon^{ijkl}g_{abc}\,\eta^{a}\left(A_{j}^{b}-\bar{A}_{j}^{b}\right)\Bigg[3\left(F_{kl}^{c}+\bar{F}_{kl}^{c}\right)-f^{c}_{de}\left(A_{k}^{d}-\bar{A}_{k}^{d}\right)\left(A_{l}^{e}-\bar{A}_{l}^{e}\right)\Bigg] \ ,
\\
&Q_{\rm diff}\left[\zeta\right]=-\frac{\kappa}{2}\int_{\partial\Sigma} d^3x\,\hat{n}_{i} \,\epsilon^{ijkl}g_{abc}\,\zeta^{m}(A^{b}_{j}-\bar{A}^{b}_{j})
\Bigg[A^{a}_{m}\left(2F^{c}_{kl}+\bar{F}^{c}_{kl}-\frac{1}{2}f^{c}_{de}(A^{d}_{k}-\bar{A}^{d}_{k})(A^{e}_{l}-\bar{A}^{e}_{l})\right)
\\
&\hskip3truecm+\bar A^{a}_{m}\left(F^{c}_{kl}+2\bar{F}^{c}_{kl}-\frac{1}{2}f^{c}_{de}(A^{d}_{k}-\bar{A}^{d}_{k})(A^{e}_{l}-\bar{A}^{e}_{l})\right)
\Bigg] \ .
\eal
\eeq
Moreover, comparing the space-time decomposition of \eqref{t4+1} and the general expressions in \eqref{csah} and \eqref{bt}, we also find
\beq
K_{a}\left(A\right) =  3\kappa\epsilon^{ijkl}g_{abc}F^b_{ij}F^c_{kl} \ ,\qquad
\rho_{a}^{ij}\left(A,\bar{A}\right)  =  -\kappa \epsilon^{ijkl}g_{abc} A_{k}^b\bar{A}_{l}^c \ ,
\eeq
which can be used to write down the symmetry generators obtained by evaluating \eqref{gauge_functional} and \eqref{diffeo_functional}. The generator of gauge transformations \eqref{gauge_functional} is given by
\beq\label{gauge_constraint_4+1}
\bal
G\left[\eta,\bar{\eta}\right] & = \kappa \int_{\Sigma}d^{4}x\Bigg\{-D_{i} \eta^{a}\pi_{a}^{i}-\bar{D}_{i}\bar\eta^{a}\bar{\pi}_{a}^{i}+\epsilon^{ijkl}g_{abc}\Bigg[\frac{3}{4}\left(\eta^{a} F_{ij}^{b}F_{kl}^{c}-\bar\eta^{a} \bar{F}_{ij}^{b}\bar{F}_{kl}^{c}\right)
\\&
+D_{i}\eta^{a}\left(F_{jk}^{b}A_{l}^{c}-\frac{1}{4}f_{de}^{c}A_{j}^{b}A_{k}^{d}A_{l}^{e}\right)
-\bar{D}_{i}\bar\eta^{a}\left(\bar{F}_{jk}^{b}\bar{A}_{l}^{c}-\frac{1}{4}f_{de}^{c}\bar{A}_{j}^{b}\bar{A}_{k}^{d}\bar{A}_{l}^{e}\right)\Bigg]\Bigg\}+Q_{\rm gauge}\left[\eta\right] \ ,
\eal
\eeq
whereas the generator of diffeomorphisms reads
\beq
\bal
L\left[\zeta,\bar{\zeta}\right] & =
\int_\Sigma   d^{4}x\Bigg\{ -\mathfrak{L}_{\zeta}A^a_i \pi^i_a
- \mathfrak{L}_{\bar\zeta}\bar A^a_i \bar\pi^i_a
+\kappa\epsilon^{ijkl}g_{abc}\Bigg[\frac{3}{4}\left( \zeta^m A^{a}_{m}F^{b}_{ij}F^{c}_{kl} -  \bar \zeta^m
\bar{A}^{a}_{m}\bar{F}^{b}_{ij}\bar{F}^{c}_{kl}\right)
\\&+
\mathfrak{L}_{\zeta}A^{a}_{i}\left(F^{b}_{jk}-\frac{1}{4}f^{b}_{de}A^{d}_{j}A^{e}_{k}\right)A^{c}_{l}
-
\mathfrak{L}_{\bar\zeta}\bar{A}^{a}_{i}\left(\bar{F}^{b}_{jk}-\frac{1}{4}f^{b}_{de}\bar{A}^{d}_{j}\bar{A}^{e}_{k}\right)\bar{A}^{c}_{l}\Bigg]\Bigg\}
+ Q_{\rm diff}\left[\zeta\right] \;.
\eal
\eeq
By means of the generalized bracket \eqref{soloviev}, one can explicitly show that $G\left[\eta,\bar{\eta}\right]$ and $L\left[\zeta,\bar{\zeta}\right]$ satisfy the Poisson algebras given in \eqref{algebra_generators}.
Thus, again, one directly obtains the Dirac algebra \eqref{algebra_gauge}, \eqref{algebra_diffeo}, after setting the constraints strongly to zero. For $\bar{A}_i^a=0$, these expressions boil down to the regularized smeared generators of gauge transformations and spatial diffeomorphisms that correspond to CS theory. The associated charges \eqref{charges_4+1} reduce to the known results that have been previously derived using the Regge-Teitelboim method \cite{Banados:1994tn,Miskovic:2005di}.

\section{Conclusions}
\label{conclusions}

We have studied the Hamiltonian formulation of TrFT in arbitrary odd space-time dimensions considering regular and generic sectors. Instead of adopting the standard approach of regularizing the relevant functionals by adding boundary terms that cancel boundary variations, we have allowed the presence of surface integrals in the variations of the constraints in order to track down the contribution of the transgression boundary term throughout Dirac's algorithm. By doing so, we have found useful the prescription by Soloviev and Bering that extends the definition of the Poisson bracket to the boundary. As shown in \Cref{section_brackets}, for a theory whose action depends only on the fields and their first derivatives, using such generalized Poisson bracket is equivalent to acting with standard Poisson brackets inside the boundary terms, allowing to retain the contribution of the surface integrals coming from partial integration. The effect of keeping all the boundary contributions produced by the generalized Poisson bracket leads to surface integrals for the generators of gauge transformations \eqref{gauge_functional} and diffeomorphisms \eqref{diffeo_functional} of the theory. We have developed in detail the $D=2+1$ and $D=4+1$ cases, where we have shown that the resulting surface integrals appearing in the symmetry generators match the ones previously found in the literature by means of Noether's theorem and can be interpreted as the charges associated with these transformations \cite{Mora:2006ka}. The generalized Poisson bracket also allows obtaining the right Poisson algebra of the charges, which do not possess central terms due to the gauge-invariant character of the action. Since setting one of the connections in the transgression to zero reduces the gauge and diffeomorphism charges to the CS case, it is not possible to obtain the known central terms in the surface algebra associated with CS theories by simply evaluating the charge algebra found in the transgression case. Instead, in the CS case, after finding the expression of the conserved charges, their algebra should be computed once again to obtain a central term. We hope to address the emergence of such central charges in the context of our formalism elsewhere.

The generalized Poisson bracket \eqref{soloviev} thus provides an alternative approach to compute conserved charges in field theory and gravity in a straightforward manner. Also, it could be advantageous as a tool in the description of asymptotic symmetries and conserved charges in gauge and gravity theories, where boundary terms defining such charges are customary \cite{Hanson_Regge_Teitelboim}. However, it is important to highlight that the exceptional structure of the transgression boundary term seems to be of primary importance for this method to produce the right final result at the level of the symmetry generators. It is, therefore, likely that for this procedure to be well-defined, the initial action of a given theory must contain the right boundary term to regularize the action after imposing some boundary conditions.

An interesting future direction is the relation between the Poisson algebra of surface charges in transgression field theory and asymptotic symmetries in gravity. A good example is the Bondi, van der Burg, Metzner and Sachs (BMS) algebra \cite{Bondi:1962px,Sachs:1962wk}. Even though Chern-Simons gravity in five dimensions is known to be different from five-dimensional General Relativity \cite{Troncoso:1998ng,Mora:2004kb,Izaurieta:2005vp,Izaurieta:2006aj,Izaurieta:2009hz}, asymptotic symmetries of this type can emerge from the structure of the transgression form. Indeed, a five-dimensional version of the BMS algebra has been recently constructed \cite{Fuentealba:2021yvo}. Given certain boundary conditions, a five-dimensional gravity theory based on a TrFT could give rise to an asymptotic symmetry of this type. Another possible future direction is the use of TrFT in the description of certain condensed matter systems. Indeed, non-relativistic Chern-Simons forms provide effective models for topological phases of matter, such as the fractional quantum Hall effect, topological insulators and topological superconductors (see, for example, \cite{bernevig2013topological}). Moreover, transgression forms have been used to define effective boundary descriptions of topological insulator interfaces \cite{Acik:2013ija}, which may allow for higher-dimensional generalizations. Along the same lines, a transgression generalization of the pure CS construction given in \cite{Salgado-Rebolledo:2021nfj} for fractional quantum Hall states could be useful in the geometric description of bi-layer Hall systems. Finally, it would be interesting to generalize our construction to the case of the gauged-WZW models considered in \cite{Anabalon:2006fj}, which can be obtained from a TrFT when one of the gauge connections is a gauge transformation of the other one. The motivation in this case is that WZW models are relevant in the edge description of higher-dimensional versions of the quantum Hall effect \cite{Karabali:2004km,Polychronakos:2004sw}.

\section*{Acknowledgments}
We wish to thank M.~Campiglia, P.~Mora, C.~Troessaert, R.~Troncoso and J.~Zanelli for very useful discussions and comments during different stages of this work. P.~P. is supported by Fondo Nacional de Desarrollo Cient\'{i}fico y Tecnol\'{o}gico--Chile (Fondecyt Grant No.~3200725) and by Charles University Research Center (UNCE/SCI/013). P.~S-R. has received funding from the Norwegian Financial Mechanism 2014-2021 via the Narodowe Centrum Nauki (NCN) POLS grant 2020/37/K/ST3/03390. A.~V. has been partially funded by the National Agency for Research and Development ANID SIA SA77210097. P.~P. and P.~S-R. acknowledge the warm hospitality of the Centro de Estudios Cient\'ificos (CECs) in Valdivia, Chile, during different stages of this work.

\bibliographystyle{utcaps}
\bibliography{Transgression_biblio}

\end{document}